\documentclass[%
reprint,
superscriptaddress,
showkeys,
bibnotes,
amsmath,amssymb,
aps,
]{revtex4-2}

\usepackage{amssymb}
\usepackage{amsmath}
\usepackage{amsthm}
\usepackage{eucal}
\usepackage{mathrsfs}
\usepackage{graphicx}% Include figure files
\usepackage{dcolumn}% Align table columns on decimal point
\usepackage{bm}% bold math
\usepackage{hyperref}% add hypertext capabilities
\usepackage[usenames,dvipsnames,x11names,table]{xcolor}

\usepackage{soul}
\usepackage{balance}
\usepackage{dsfont}
\usepackage{tikz}
\usetikzlibrary{shapes}
\usepackage{multirow}
\usepackage{tabularx}
\usepackage{enumitem}
\usepackage{cancel,soul}
\setlist[itemize]{align=parleft,left=0pt..1em}
\usepackage{color}

\newcommand{\orangecircle}{\raisebox{0.5pt}{\tikz{\node[draw=black,scale=0.6,circle,fill=orange](){};}}}

\newcommand{\blackcircle}{\raisebox{0.5pt}{\tikz{\node[draw=black,scale=0.6,circle,fill=black](){};}}}

\definecolor{darkslategray}{HTML}{2F4F4F}
\newcommand{\tealcross}{\protect\tikz[]{\protect\draw[darkslategray, scale=0.5] (0,0) -- (0.4,0.4) (0,0.4) -- (0.4,0);}}

\newcommand{\redtriangle}{\protect\tikz[baseline=-0.2ex]{\protect\draw[fill=red,scale=0.5,draw=black] (0,0) -- (0.2,0.4) -- (0.4,0) -- cycle;}}

\newcommand{\orangetriangle}{\protect\tikz[]{\protect\draw[fill=orange!50,scale=0.6,draw=black] (0,0) -- (0.15,-0.3) -- (0.3,0) -- cycle;}}

\newcommand{\Ra}{\operatorname{Ra}}
\newcommand{\Nu}{\operatorname{Nu}}

\makeatletter
\let\@internalcite\cite
\def\cite{\def\citeauthoryear##1##2{##1, ##2}\@internalcite}
\def\shortcite{\def\citeauthoryear##1##2{##2}\@internalcite}
\def\@biblabel#1{\def\citeauthoryear##1##2{##1, ##2}[#1]\hfill}
\makeatother

\begin{document}

\preprint{APS/123-QED}

\title{Intermittent thermal convection in jammed emulsions}

\author{Francesca Pelusi}
\email{francesca.pelusi@cnr.it}
\affiliation{Istituto per le Applicazioni del Calcolo, CNR - Via Pietro Castellino 111, 80131 Naples, Italy}
\author{Andrea Scagliarini}
\affiliation{Istituto per le Applicazioni del Calcolo, CNR - Via dei Taurini 19, 00185 Rome, Italy}
\affiliation{INFN, Sezione Roma "Tor Vergata", Via della Ricerca Scientifica 1, 00133 Rome, Italy}
\author{Mauro Sbragaglia}
\affiliation{Department of Physics \& INFN, University of Rome  Tor Vergata, Via della Ricerca Scientifica 1, 00133 Rome, Italy}
\author{Massimo Bernaschi}
\affiliation{Istituto per le Applicazioni del Calcolo, CNR - Via dei Taurini 19, 00185 Rome, Italy}
\author{Roberto Benzi}
\affiliation{Department of Physics \& INFN,  University of Rome Tor Vergata, Via della Ricerca Scientifica 1, 00133 Rome, Italy}
%\vspace{0.5cm}
\date{\today}
%%%%%%%%%%%%%%%%
\begin{abstract}
%%%%%%%%%%%%%%%%

\noindent We study the process of thermal convection in jammed emulsions with a yield-stress rheology. We find that heat transfer occurs via an intermittent mechanism, whereby intense
short-lived convective ``heat bursts'' are spaced out by long-lasting conductive periods. 
This behaviour is the result of a sequence of fluidization-rigidity transitions, 
rooted in a non-trivial interplay between emulsion yield-stress rheology and plastic activity, which
we characterize via a statistical analysis of the dynamics at the droplet scale.
We also show that droplets' coalescence induced during heat bursts leads to a spatially heterogeneous phase-inversion of the emulsion which eventually supports a sustained convective state.
\end{abstract}

\maketitle
%%%%%%%%%%%%%%%%%%%%%%%
\section{Introduction}
%%%%%%%%%%%%%%%%%%%%%%%
Concentrated emulsions are binary mixtures of immiscible liquids (say ``oil" and ``water"), stabilized against phase separation by the presence of surfactants, where droplets of a given phase (e.g., oil) are dispersed in a continuous phase (e.g., water, thus making an oil-in-water, or, in short, O/W, emulsion). When the volume fraction of the dispersed phase is large enough, droplets are generally strongly deformed, and the emulsion displays a complex non-Newtonian rheology with a non-vanishing yield stress, which is peculiar of materials referred as yield stress materials (YSMs)~\cite{Larson,Coussot05,Bonn17}. These materials have been extensively characterized from the rheological point of view. Still, much less is known about their behavior in buoyancy-driven thermal flows, despite their relevance particularly for geophysical problems, such as mantle convection~\cite{Orowan65,Morgan71,Montelli06,French15,Davaille2018} and lava flows~\cite{Griffiths2000} or for analog modeling of geodynamic processes~\cite{SchellartStrak,DiGiuseppe2014,Reber2020}, 
but also in pharmaceutical~\cite{MartiMestres02} and food cooking~\cite{Jeffreys57} applications. The study of YSMs under thermal convection also takes on a great interest from a fundamental point of view: unlike more standard shear or pressure-driven flows, the forcing is local and dynamic, so the stress configuration across the system is not given {\it a priori}, but the dynamics itself instead determine it. The non-trivial correlation between such stress configuration and the complex structural fragility landscape can lead to unexpected phenomena in a disordered medium.
%%%%%%%%%%%%%%%%%%%%%%%%%%%%%%%%%%%%%%%%%%%%%%%%%%%%%%%%%%%
\begin{figure*}[t!]
\centering
\includegraphics[width=1.\linewidth]{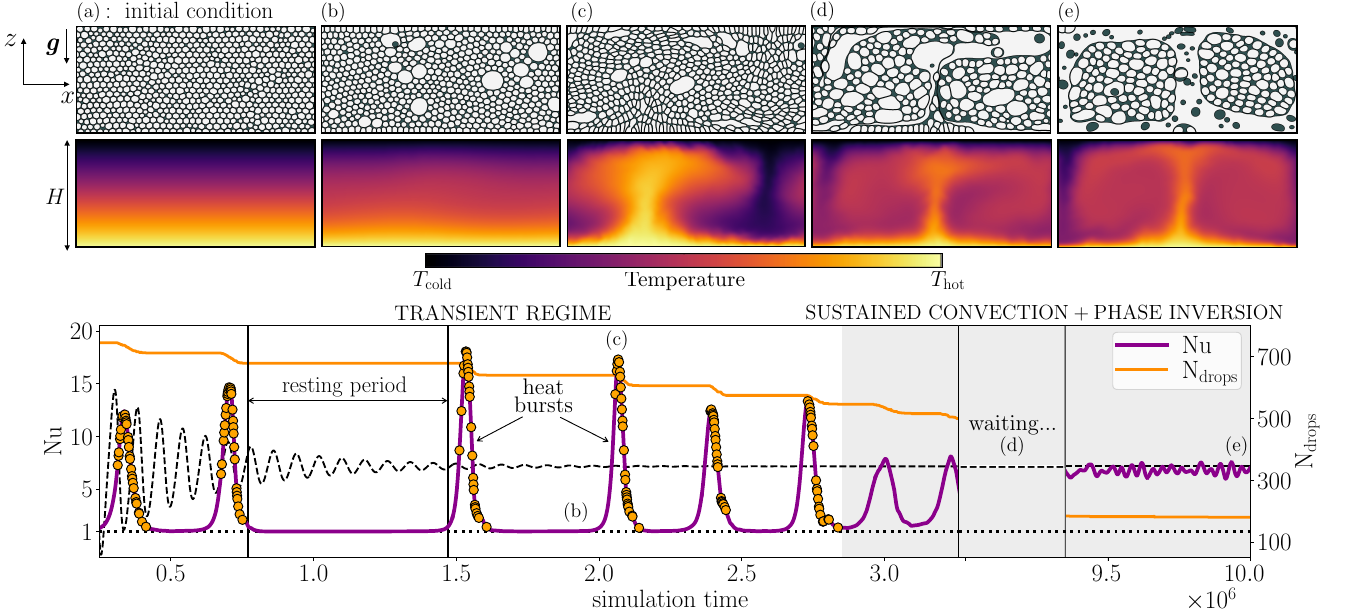}
\caption{Dynamics of a jammed emulsion in a Rayleigh-B\'enard setup: the time evolution of the average heat transfer is quantified via the Nusselt number $\Nu$, shown together with the number of droplets N$_{\mathrm{\tiny drops}}$. Simulation time is reported in lattice Boltzmann units. A decrease in N$_{\mathrm{\tiny drops}}$ corresponds to the occurrence of coalescence events (an orange bullet (\protect \orangecircle) is drawn any time at least a coalescence event occurs). The dashed black line marks the time evolution of $\Nu$ for a homogeneous system under the same conditions, while the dotted black line highlights the conductive regime ($\Nu = 1$). Upper panels show maps of density (first row) and temperature fields (second row) of selected instants of time ((a)-(e)).}\label{fig:sketch}%
\end{figure*}
%%%%%%%%%%%%%%%%%%%%%%%%%%%%%%%%%%%%%%%%%%%%%%%%%%%%%%%%%%
Experimental~\cite{Davailleetal13,Darbouli13,Kebicheetal14,Metivier17,Jadhav21}, theoretical and numerical studies~\cite{Turanetal12,Karimfazli16,Li16,Aghighi18,Masoumi19,Santos21,Aghighi23} focused on the onset of convection, looking at the steady-state value of the heat flux (or, conversely, of the temperature difference, in experiments with controlled heat flux) drawing an overall picture whereby yield stress rheology hinders convection. However, the character, the way it is approached, and the very existence of a convective steady-state deserve a in-depth discussion, and this has to be considered an open problem. Experiments with microgels evidenced how the development of a thermal plume is associated with structural damages in the material, suggesting that plastic microdynamics might play a role. Moreover, unlike in Newtonian convection, the plume can stop and cool down until the emergence of a second plume~\cite{Davailleetal13}. Convection arrest following a long period of chaotic oscillations 
was observed also in numerical simulations~\cite{Vikhansky09}. However, a comprehensive study of the long-time behavior of thermal convection in YSMs and the interplay between microscopic constituents and non-linear rheology is still missing. In this Letter, we show that convective heat transfer in a YSM occurs via intermittent heat flux bursts associated with fluidization of the system (see Fig.~\ref{fig:sketch}). The origin of such complex dynamics resides in the elastoplastic character of the material and, therefore, its explanation requires a description that takes into account its granularity (the finite-size droplets microdynamics in the case of emulsions), as suggested also in a previous experimental work~\cite{Jadhav21}. In particular, in concentrated emulsions, droplet coalescence eventually leads to a spatially heterogeneous phase inversion, which supports a sustained convective state (see Fig.~\ref{fig:sketch}(e)).\\
We perform numerical simulations of a two-dimensional jammed O/W emulsion in the Rayleigh-B\'enard setup, i.e., between two parallel no-slip walls at a distance $H$ with a temperature difference $\Delta T = T_{\text{hot}} - T_{\text{cold}}$, subject to the action of gravity (with strength $g$) in the direction of the thermal gradient (see Fig.~\ref{fig:sketch}(a))~\cite{Grossmann01,Ahlers09,Chilla12}. The Rayleigh number is  $\Ra = \alpha g \Delta T H^3/(\kappa \nu) \approx 4 \times 10^5$, where $\alpha$ is the thermal expansion coefficient, $\kappa$ the thermal diffusivity, and $\nu$ the kinematic viscosity, while the Prandtl number Pr$=\nu/\kappa$ is fixed to unity. Notice that, we define $\Ra$ and Pr in terms of the kinematic viscosity of the homogeneous liquid phase (either the carrier phase or the dispersed phase, which are equiviscous) since the emulsion structure, and, in turn, rheology changes over time. Further details on the employed methodology are reported in the Supplementary Material (SM)~\footnote{See Supplemental Material, which includes Refs.~\cite{Benzi92,Kruger17,Succi18,Lohse10,Benzi09,Sbragagliaetal12,Dollet15,Gross2014,sun2016three,aouane2021structure,wouters2021lattice,guglietta2023suspensions}, for additional information about the numerical methodology.}.
%%%%%%%%%%%%%%%%%%%%%%%%%%%%%%%%%%%%%%%%%%%%%%%%%%%%%%%%%%%%%%%%%%%%%%%%%
\section{Macroscopic Analysis}
%%%%%%%%%%%%%%%%%%%%%%%%%%%%%%%%%%%%%%%%%%%%%%%%%%%%%%%%%%%%%%%%%%%%%%%%%
After the emulsion preparation (see SM~[31]), and the application of a finite perturbation to the hydrodynamical velocity to trigger the thermal convection, we probe the time response of the system by instantaneously measuring the vertical velocity $u_z=u_z(x,z,t)$ and the temperature $T=T(x,z,t)$ and computing the dimensionless heat flux expressed by the Nusselt number 
\begin{equation}\label{eq:Nusselt}
\Nu = 1 + \frac{\langle u_z T \rangle_{x,z}}{\kappa \frac{\Delta T}{H}}\ ,
\end{equation}
where $\langle \dots \rangle_{x,z}$ denotes the space average~\cite{Shraiman90,Ahlers09,Verzicco10,Chilla12}. For a single phase (homogeneous) fluid in a purely conductive state, i.e., below the critical value for convection $\Ra_c$~\cite{Chandrasekhar61}, the advective contribution $\langle u_z T \rangle_{x,z}$ is zero and $\Nu = 1$ (dotted black line in Fig.~\ref{fig:sketch}). As $\Ra$ is increased above $\Ra_c$, the homogeneous fluid reaches a state of steady convection with a large-scale flow, which increases the heat flux in regard to the conduction ($\Nu >1$). Further, for sufficiently large values of Ra, but ahead of the transition to turbulence, there exist stationary thermal convective states that are stable despite showing oscillations whose amplitude, however, decays exponentially over time (dashed black line in Fig.~\ref{fig:sketch})~\cite{Ecke86}. This scenario changes in the case of emulsions, where the value of $\Ra_c$ and the onset to thermal convection depend on the volume fraction $\phi$ of the initially dispersed phase. It has been shown in Refs.~\cite{PelusiSM21,PelusiSM23} that the presence of droplet interactions induces fluctuations in the heat flux (i.e., fluctuations of $\Nu$), which increase with $\phi$. Following those works, a natural question arises regarding the behavior of emulsions in thermal convection when $\phi$ is large enough to make the emulsion exhibiting a yield stress: droplets get strongly deformed and endow the emulsion with complex elastic-viscoplastic properties typical of YSMs~\cite{PrincenKiss89,Zhang06,Balmforthetal14,Bonn17}. Hence, we consider jammed emulsions~\footnote{The volume fraction of the initially dispersed phase is $\phi=80\%$, corresponding to an effective volume fraction $\phi_{\text{eff}} \approx \phi (1 + 2h/d) \approx 93\%$, corrected as in~\cite{MBWprl1995} to account for the film thickness $h$} that exhibit a finite yield stress (see SM~\footnote{See Supplemental Material, which includes Ref.~\cite{Mason96}, for a detailed rheological characterization.}). In Fig.~\ref{fig:sketch}, we report the time evolution of $\Nu$ for the emulsion (solid purple line) along with some snapshots of the corresponding density and temperature fields at selected times (panels (a)-(e)). The phenomenology is remarkably different from what observed in Refs.~\cite{PelusiSM21,PelusiSM23} at lower values of $\phi$: in the jammed state (i.e., as long as the emulsion can be considered a YSM), the system exhibits a dynamics characterized by extreme heat transfer bursts (convective periods), with $\Nu \gg 1$, spaced out by long-lasting resting (conductive) periods, with $\Nu \approx 1$. During the duration of the heat bursts, droplets' coalescence takes place (orange bullets), progressively reducing the number of droplets (N$_{\mathrm{\tiny drops}}$, solid orange line). Coalescence events decrease the interface energy and eventually lead to the inversion from an O/W to a W/O emulsion, causing the system to persist in a convective state. However, due to the self-organized flow structures (and associated stress configuration), the phase inversion is not global in space but takes place heterogeneously, with patches of concentrated O/W emulsion around the center of the convective rolls in a matrix of phase-inverted, diluted, W/O emulsions (see Fig.~\ref{fig:sketch}(e)). The characterization of the phase-inverted state is particularly intriguing as its rheology diverges significantly from the jammed O/W emulsion (see SM, Fig.2 [42]), warranting a dedicated investigation. Here, we focus on the unexpected dynamics in the jammed emulsion. 
In this regime, in correspondence with $\Nu$ peaks, the material fluidizes, and convection mixes the temperature field, causing variations of the stress whose amplitude can, locally, go below the yielding point, conferring rigidity again to the material. Surprisingly, after the emulsion stops resting, the motion can restart with a heat transfer burst before the system stops again. The sequence of heat bursts does not display the same regularity of the oscillations in $\Nu$ that are observed for the homogeneous system (dashed black line in Fig.~\ref{fig:sketch}), and the resulting time needed to achieve the sustained convective state is much longer than the time it takes for a homogeneous system (see movie.mp4). The portrayed scenario is far from trivial, highlighting that the transition to thermal convection of a highly packed emulsion made up of soft domains shows an anomalous intermittent behavior that profoundly delays the transition to a sustained convective state.
This intermittency persists for a finite range of values of $\Ra$ below which convection is suppressed and above which phase inversion is triggered without intermittent transient (see SM, Fig.5). We remark that the observed intermittent behavior arises from the non-zero yield stress in the examined jammed emulsion. Attempts to observe intermittency in systems without yield stress failed (see SM, Figs.3 and 4).
Furthermore, notice that different initial configurations lead to a different duration of the transient intermittent regime and number of heat bursts (see SM, Fig.1 [31]).
\\
We remark that any attempt to understand this intermittency in terms of continuum equations equipped with a local constitutive relation between stress and shear rate, regularized as the stress tends to the yield value~\cite{Turanetal12,BalmforthRust09,Masoumi19,Zhang06}, would unavoidably fail. Once the system stops, no mechanism can set it back in motion. Such an approach, indeed, provides misleading information since it is not able to capture the role played by the presence of finite-sized droplets and the change in local microstructure with plastic rearrangements~\cite{Goyon08}, 
that represent the source of intrinsic effective ``noise"  able to trigger convection again. Therefore, systematically characterizing the observed scenario would require analysis at scales comparable with the droplet size.
%%%%%%%%%%%%%%%%%%%%%%%%%%%%%%%%%%%%%%%%%%%%%%%%%%%%%%%%%%%%%%%%%%%%%%%%%
\section{Microscopic Analysis}
\begin{figure}[t!]
\centering
\includegraphics[width=1.\linewidth]{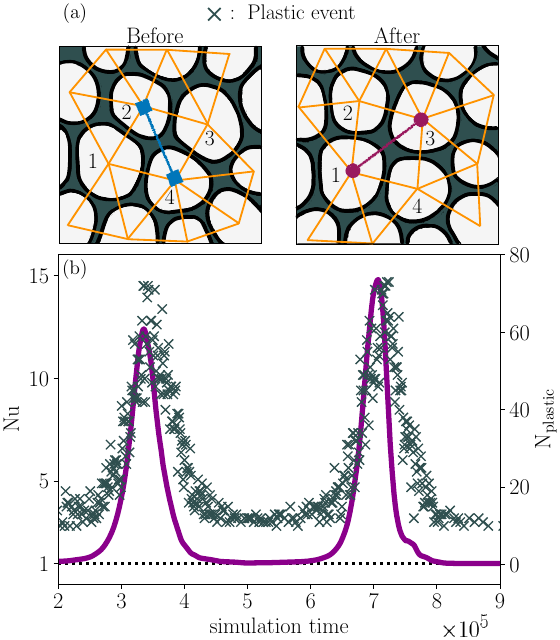}
\caption{Panel (a): a sketch describing the definition of plastic event starting from the Delaunay triangulation (orange links). A plastic event takes place when a neighboring droplets switching occurs, corresponding to the disappearance of the link between two droplets (2 and 4) and the creation of a new link between other two (1 and 3). Panel (b): Nusselt number Nu (purple solid line) and the number of occurred plastic events N$_{\mathrm{\tiny plastic}}$ (\tealcross) as a function of the simulation time, expressed in lattice Boltzmann units.}\label{fig:plastic}%
\end{figure}

%%%%%%%%%%%%%%%%%%%%%%%%%%%%%%%%%%%%%%%%%%%%%%%%%%%%%%%%%%%%%%%%%%%%%%%%%
Next, we focus on droplets' microstructure and droplet contribution to heat transfer. In Fig.~\ref{fig:plastic}, we report the detection of changes in droplet microstructures via the analysis of the Delaunay triangulation constructed from the droplets center of mass~\cite{Bernaschi16}. Through the Delaunay triangulation, a plastic rearrangement is identified whenever one edge of a droplet shrinks to zero, and neighboring droplets switching occurs (see Fig.~\ref{fig:plastic}(a)). In Fig.~\ref{fig:plastic}(b), we report signal of the Nusselt number $\Nu$ along with the number of plastic rearrangements N$_{\mbox{\tiny plastic}}$ in a representative time interval where both heat bursts and resting periods are present. As expected, N$_{\mbox{\tiny plastic}}$ grows when the emulsion enters a convective state (heat bursts, $\Nu \gg 1$) but, more interestingly, a residual plastic activity is also apparent during the resting periods, revealing that the macroscopic conductive periods possess non-trivial features at the level of droplets' microstructure. 
%%%%%%%%%%%%%%%%%%%%%%%%%%%%%%%%%%%%%%%%%%%%%%%%%%%%%%%%%%%%%%%%%%%%%%%%%%%%%%%%%%%%%%%%%%%%%%%%%%%%
%%%%%%%%%%%%%%%%%%%%%%%%%%%%%%%%%%%%%%%%%%%%%%%%%%%%%%%%%%%%%%%%%%%%%%%%%%%%%%%%%%%%%%%%%%%%%%%%%%%%%
The latter evidence stimulated further analysis to characterize the heat transfer at the droplet scale. Starting from the $i-$th droplet and the associated center-of-mass location $\mathbf{X}_i(t)= (X_i(t),Z_i(t))$ at a generic time $t$, we considered the velocity, $u^{(i)}_z=u_z(\mathbf{X}_i(t),t)$, temperature, $T^{(i)}=T(\mathbf{X}_i(t),t)$, and temperature gradient, $(\partial_z\,T)^{(i)}=\partial_z T(\mathbf{X}_i(t),t)$, and we defined the Nusselt number at the droplet scale $\Nu_{i}^{(\mbox{\tiny drop})}$~\cite{PelusiSM21,PelusiSM23}
\begin{equation}\label{eq:NuDrops}
\Nu^{(\mbox{\tiny drop})}_{i} = \frac{u^{(i)}_z  T^{(i)} - \kappa (\partial_z\,T)^{(i)}}{\kappa \frac{\Delta T}{H}}.
\end{equation}
%%%%%%%%%%%%%%%%%%%%%%%%%%%%%%%%%%%%%%%%%%%%%%%%%%%%%%%%%%%%%%%%%%%%%%%%%%%%%%%%%%%%%%%%%%
\begin{figure}[t!]
\centering
\includegraphics[width=1.
\linewidth]{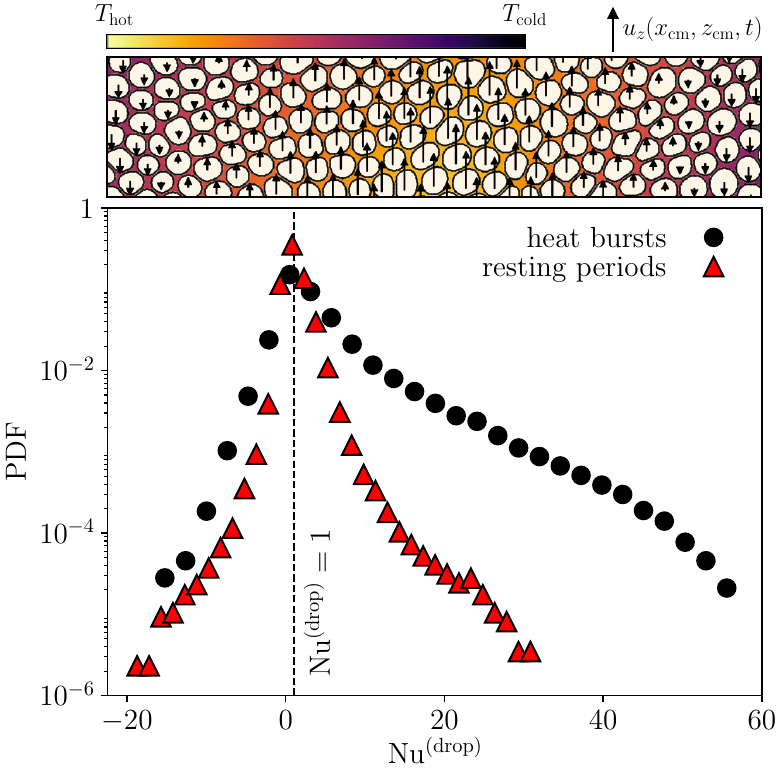}
\caption{Probability distribution functions (PDFs) of the Nusselt number at the droplet scale $\Nu^{(\mathrm{\tiny drop})}$ (see Eq.~\eqref{eq:NuDrops}), for droplets contributing to heat bursts (\protect \blackcircle) and resting periods (\redtriangle). Upper panel: temperature map along with the droplets' center-of-mass, with the corresponding vertical velocity vectors used to compute  $\Nu^{(\mathrm{\tiny drop})}$.} \label{fig:PDF_nuDrop}%
\end{figure}
%%%%%%%%%%%%%%%%%%%%%%%%%%%%%%%%%%%%%%%%%%%%%%%%%%%%%%%%%%%%%%%%%%%%%%%%%%%%%%%%%%%%%%%%
%%%%%%%%%%%%%%%%%%%%%%%%%%%%%%%%%%%%%%%%%%%%%%%%%%%%%%
\begin{figure}[t!]
\centering
\includegraphics[width=1.
\linewidth]{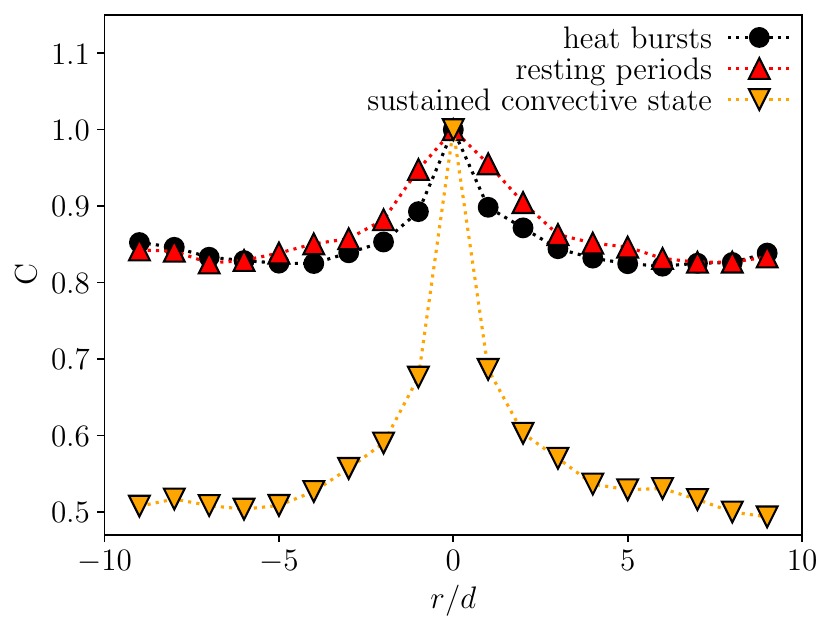}
\caption{Correlation function C  of interface displacement (see Eq.~\eqref{eq:correlation}) as a function of the distance from the center of the channel $r$ normalized with the average droplet diameter $d$. We compare correlation functions for heat bursts (\protect \blackcircle) and resting periods (\redtriangle). Correlation function computed during the sustained convective state after the transient dynamics is also shown (\orangetriangle).}\label{fig:corr_func}%
\end{figure}
%%%%%%%%%%%%%%%%%%%%%%%%%%%%%%%%%%%%%%%%%%%%%%%%%%%%%%%%%%%%%%%%%%%%%%%%%
This definition is such that the macroscopic Nusselt number $\Nu$ (Eq.~\ref{eq:Nusselt}) can be reconstructed as the sum over the contributions of each single droplet, $\Nu \approx \frac{1}{\mathrm{N}_{\mbox{\tiny drops}}}\sum_{i=1}^{\mathrm{N}_{\mbox{\tiny drops}}} \mbox{Nu}^{(\mbox{\tiny drop})}_{i}$ (see SM, Fig.~6). Via the analysis of $\Nu_{i}^{(\mbox{\tiny drop})}$, we can shed light on the microscopic origin of the intermittency in $\Nu$ observed at macroscales. Fig.~\ref{fig:PDF_nuDrop} shows the probability distribution function (PDF) for $\Nu^{(\mbox{\tiny drop})}$ during the transient intermittent dynamics, where we separately examined the contributions to heat bursts periods and the resting periods. Specifically, we collected the values of $\Nu_{i}^{(\mbox{\tiny drop})}$ associated with every droplet at any time during the two periods mentioned above and, to improve the statistics, we considered tens of numerical simulations for different initial configurations. The resulting statistics of $\Nu^{(\mbox{\tiny drop})}$ reveals enhanced tails and clear intermittency during the heat bursts. In fact, the average value of the distribution is sensibly larger than the most probable value; additionally, when the system is at rest, the statistics of $\Nu^{(\mbox{\tiny drop})}$ features a non-zero variance, indicating that heat is transferred at the scales of the droplets, despite, on average, the system is in a conductive state at the macroscopic scale ($\Nu \approx 1$). The presence of residual elastoplastic activity shown both in Fig.~\ref{fig:plastic} and in Fig.~\ref{fig:PDF_nuDrop} is inherently related to the presence of finite-size droplets and acts as a perturbation on the system during the macroscopic resting periods. A natural question arises on the mechanism that allows the system to move from the resting periods to the heat bursts. The effect of localized perturbations can be spread over the system size if its spatio-temporal correlation is sufficiently large~\cite{PelusiAvalanche19}. The presence of residual plastic activity due to spatial heterogeneity may catalyze the observed irregularity and intermittency in heat transfer bursts. A confirmation of the system's extensive spatial correlation is, therefore, essential. To define a quantitative measure of the correlations associated with the motion of the interface during and after the transient dynamics (i.e., when it becomes difficult to identify and follow each droplet), we follow the procedure introduced in Ref.~\cite{Benzietal16}: we divide the domain into squares with a side of length $d$ (average droplet diameter), then for each square $k$ we consider two consecutive times $t$ and $t+\tau$. We compute the interface displacement as 
$A^{(\tau)}(x_k,z_k,t) =\langle \left[\rho_O(x,z,t+\tau)-\rho_O (x,z,t)\right]^2 \rangle_{k}$, 
where $\rho_O$ is the density field of the ``oil'' phase and $\langle \dots \rangle_{k}$ denotes the average in square $k$ with center coordinates $(x_k,z_k)$. Then, we compute the correlation function C of $A^{(\tau)}_r = A^{(\tau)}(x_k,r,t)$ as
\begin{equation}\label{eq:correlation}
\mbox{C} = \left\langle\dfrac{\langle A^{(\tau)}_0 A^{(\tau)}_r \rangle_t- \langle A^{(\tau)}_0\rangle_t \ \langle A^{(\tau)}_r \rangle_t}{\sigma_0 \sigma_r} \right\rangle_x \ ,
\end{equation}
where $r$ denotes the distance from the channel center ($z_k = 0$) along the $z$-direction, $\langle \dots \rangle_{t}$ is the average over times belonging to the three different regimes (see SM), and $\sigma_r$ is the standard deviation of $A^{(\tau)}_r$. C is reported in Fig.~\ref{fig:corr_func}. It emerges that, during heat bursts, the thermal plume dynamically engages nearly the entire system, resulting in a notable correlation. Surprisingly, this correlation remains large even during resting periods, which is unexpected in a homogeneous system in a conductive state, where the absence of fluid motion typically implies no motion, hence the lack of any correlation. Notably, when the system sustains a convective state and initiates phase inversion, the correlation decreases, evidencing the absence of intermittent behavior.
%%%%%%%%%%%%%%%%%%%%%%%%%%%%%%%%%%%%%%%%%%%%%%%%
\section{Conclusions}
%%%%%%%%%%%%%%%%%%%%%%%%%%%%%%%%%%%%%%%%%%%%%%%%%%%%%%%%%%%%%%%%%%%%%%
We studied the long-time behavior of a typical YSM, i.e., a jammed emulsion, under thermal convection. We pointed out a novel phenomenon, never reported before, whereby convection occurs through fluidization-rigidity transitions, entailing heat flux bursts occurring in an otherwise conductive background. Once thermal convection is initiated, temperature field mixing causes stress fluctuations leading to a halting of convection if the stress drops below the yield value. Leveraging on numerical simulations~\cite{TLBfind22}, we can observe that, remarkably, residual plastic activity persists even during macroscopic resting periods, thus acting as a latent trigger for subsequent convective events. This plasticity stems from the inherently disordered structure of the emulsion, resulting in an intermittent onset of convective transport without discernible regularity, which cannot be predicted based on continuum equations coupled with bulk rheology. Fluidization events and heat bursts are associated with droplets' coalescence, eventually leading to a phase-inverted emulsion exhibiting spatial heterogeneity: to our knowledge, such a final state remains unexplored and warrants further investigation, both rheologically and morphologically. For a given emulsion, the intermittency phenomenon could survive in a finite range of values of the Rayleigh number $\Ra$. Indeed, for values of $\Ra$ below this range, convection is suppressed, whereas, for larger values, phase inversion is observed without transient intermittency.\\
Altogether, we argue that our findings open new fundamental views on the transition to sustained convection in jammed emulsions, which occurs through a rather unusual transient and intermittent behavior and leads to heterogeneous phase inversion. 

\section*{Acknowledgements}
This work has been carried out within the TEXTAROSSA project (G.A. H2020-JTI-EuroHPC-2019-1 No. 956831). FP, MS and MB acknowledge the support of the National Center for HPC, Big Data and Quantum Computing, Project CN\_00000013 – CUP E83C22003230001 and CUP B93C22000620006, Mission 4 Component 2 Investment 1.4, funded by the European Union – NextGenerationEU.

\bibliographystyle{apsrev4-2}
\bibliography{francesca}

\end{document}